\documentclass[twocolumn, numberedappendix]{aastex62}

\usepackage{amstext}
\usepackage{amsmath}
\usepackage{amssymb}
\usepackage{graphicx}
\usepackage{upgreek}
\usepackage{lineno}
\graphicspath{{./}{figures/}}

\def\vphi{v_{\phi}}
\def\vr{v_{\rm R}}
\def\vk{v_{\rm K}}
\def\alphak{\alpha_{\rm k}}

\begin{document}

\title{Disk draining in LIGO progenitor black hole binaries and its significance to  electromagnetic counterparts}

\correspondingauthor{Xiaoshan Huang}
\email{xshuang@caltech.edu}

\author[0000-0003-2868-489X]{Xiaoshan Huang}
\affiliation{California Institute of Technology, Astronomy Department, Pasadena, CA 91125, USA}

\author[0000-0002-3696-8035]{Sierra Dodd}
\affiliation{Department of Astronomy and Astrophysics, University of California, Santa Cruz, CA 95064, USA}

\author[0000-0003-1735-8263]{Sophie Lund Schr\o der}
\affiliation{Institute for Advanced Study, School of Natural Sciences, Princeton, NJ 08540, USA}

\author[0000-0001-7488-4468]{Shane W. Davis}
\affiliation{Department of Astronomy, University of Virginia, Charlottesville, VA 22904, USA}
\affiliation{Virginia Institute for Theoretical Astronomy, University of Virginia, Charlottesville, VA 22904, USA}

\author[0000-0003-2558-3102]{Enrico Ramirez-Ruiz}
\affiliation{Department of Astronomy and Astrophysics, University of California, Santa Cruz, CA 95064, USA}

\begin{abstract}
The effect of tidal forces on transport within a relic accretion disk in binary black holes is studied here with a suite of two-dimensional hydrodynamic simulations. As the binary contracts due to the emission of gravitational waves, the accretion disk is truncated, and a two-armed spiral wave is excited, which remains stationary in the rotating reference frame of the coalescing binary. Such spiral waves lead to increased transport of mass and angular momentum.  Our findings suggest that even in the case of weakly ionized accretion disks, spiral density waves will drain the disk long before the orbit of the two black holes decays enough for them to merge, thus dimming prospects for a detectable electromagnetic counterpart. 
\end{abstract}

\section{Introduction}
We are hailing this as the age of multi-messenger astronomy because we can now learn about remote relativistic binaries via two messengers: light and gravitational waves. As of today, the binary neutron star merger GW170817 is the only event deciphered in both messengers  \citep{2017PhRvL.119p1101A,2017ApJ...848L..13A,2017Sci...358.1556C,2017Sci...358.1559K,2017Sci...358.1579H}. Yet, the vast majority of confirmed detection by the advanced Laser Interferometer Gravitational-Wave Observatory (LIGO) and advanced Virgo observatories on Earth involve the merger between two black holes,  which, at first sight, are expected to leave no visible signal on the sky. 
The possibility that a relic accretion disk might remain at the time of the merger  \citep{2016ApJ...821L..18P,2019ApJ...875...49P, 2017ApJ...839L...7D,2016ApJ...822L...9M,2016ApJ...823L..29K,2018MNRAS.480.4732M} offers the alluring prospect that some binary black hole mergers and in particular those assembled via isolated binary evolution may produce an associated electromagnetic counterpart. While largely unsubstantiated, this 
scenario was reinvigorated by the claimed association of the binary black hole merger GW150914 with a $\gamma$-ray burst 
\citep{2018ApJ...853L...9C,2016ApJ...826L...6C}. Although we note that this association has been contested, it is thus worth exploring the possibility that certain black hole mergers could take place within a relic circumbinary disk \citep{2017ApJ...839L...7D}  or with individual accretion disks around each black hole \citep{2018ApJ...862L...3S}.

The effects of binary tidal forces on transport within a relic accretion disk are studied here with a time-dependent two-dimensional hydrodynamical model. As the black hole binary shrinks due to the emission of gravitational waves, the accretion disk is effectively truncated to radii of about half
the average radius of the Roche lobe \citep{paczynski1977model, blondin2000tidally}. A two-armed spiral wave is excited and remains stationary in the rotating reference frame of the merging binary. Such spiral waves should be a robust feature of accretion disks in binary black hole systems, and any dissipation at the spiral shock would undoubtedly transport mass and angular momentum through the disk \citep{spruit1987stationary,sawada1987standard,makita2000two,boffin2001spiral,2020ApJ...904...90P}. 

The goal of this {\it Letter}
is to provide a quantitative measure of the angular momentum transport due to spiral shocks excited in accretion disks by
the tidal forces of a black hole (BH) binary companion. To observe an electromagnetic
signal, there has to be mass remaining in the vicinity of the black holes at the time of merger. As such, if spiral shocks can effectively transport mass and angular momentum through the disk, the merger between two black holes is expected to leave no visible signature on the sky. 

\begin{figure*}[t]
    \centering
    \includegraphics[width=0.9\linewidth]{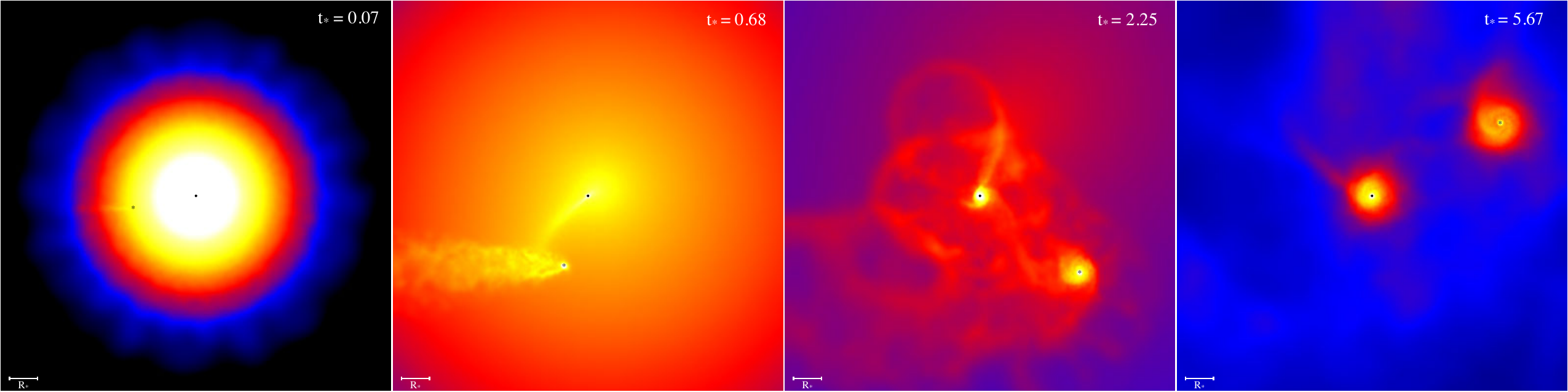}\\
    \caption{Disk formation preceding the formation of the second BH in a LIGO binary. The panels show the gas column density in the orbital plane during the first orbit of the binary after the supernova explosion. The plots are centered on the BH remnant from the exploding star, and the time stamps are given in units of the dynamical timescale of the stellar progenitor: $t_\ast =0.07,0.68,2.25,5.67.$ (from left to right). The length scale shown in all panels corresponds to the initial size of the pre-supernova progenitor: $R_\ast$. These simulations are used to inform the conditions of the gas in and around LIGO BH binaries near the time of merger. \\
    }
    \label{fig:SN_time_evo}
\end{figure*}

\section{Accretion Disk Assembly from Fallback Supernova in LIGO Binary Progenitors}\label{sec:fallbackSN}
In some binaries, the presence of the companion star has no noticeable effect, such that the evolution of the stellar members can be described independently. In most binaries, however, a variety of interactions can occur between the stars, opening up a diversity of new evolutionary pathways compared to single stellar evolution. This is particularly important for the assembly of LIGO binary progenitors, which requires BH formation as well as efficient angular momentum transport via mass transfer. This contraction is necessary to tighten the orbit so that the BH binary can merge via gravitational-wave radiation within the age of the Universe \citep{1991ApJ...380L..17P}. In this classical picture, before the collapse of the second BH, the progenitor is an isolated stellar field binary that is comprised of a Helium star with a mass $M_\star$ and the primary BH with a mass $M_1$ \citep{2016Natur.534..512B}.  
The uncertainties in the binary BH formation arise from the lack of understanding of the accretion onto the first-born BH, including the specific orbital angular momentum of the ejecta during the mass transfer phase triggered by the late-time evolution of the secondary star. A critical juncture in the life of a BH binary, regardless of the specific evolutionary pathway, is the period just after the secondary massive star explodes to leave behind a BH.  While some of the star's outer layers are likely to be ejected from the binary system, the vast majority of the inner layers eventually fall back. Depending on the properties of the stellar explosion and the binary, this fallback material can settle into accretion disks around both BHs. Accretion disk formation in BH binaries is an inherently challenging problem where the dynamics of the orbits and the hydrodynamics of the fallback material need to be taken into account consistently \citep{2017ApJ...846L..15B,2018MNRAS.476.2366F,2020A&A...635A..97B}. This final mass redistribution event provides a natural mechanism for creating a gas reservoir in and around the BH binary before merger. The fate of this relic disk around the binary depends sensitively on the angular momentum transport properties. 

Here, we follow the numerical setup described in  \citet{2017ApJ...846L..15B} and \citet{2018ApJ...862L...3S} to study the evolution of the LIGO progenitor binary system after the birth of the second BH. We make use of a modified version of the three-dimensional smoothed particle hydrodynamics (SPH) code GADGET2 \citep{2005MNRAS.364.1105S}. Our initial setup consists of $10^6$ particles with densities selected to match the profile of the 35OC KEPLER model calculated by \citet{2006ApJ...637..914W} for a $M_\star=28 M_\odot$ pre-supernova helium star with $R_\ast=0.76 R_\odot$. This model has a zero-age main sequence mass and metallicity of $M_{\rm zams}=35 M_\odot$ and $Z=0.1Z_\odot$. We assume that the innermost $3M_\odot$ of the star collapses directly into a BH, which we subsequently treat as a sink particle. To model the supernova explosion, we use a spherically symmetric kinetic piston and inject $1.15 \times 10^{52}$ erg, which corresponds to about half of the binding energy of the stellar progenitor \citep{2018ApJ...862L...3S}. 
This energy is  deposited instantaneously in a 1.5$M_\odot$ mass shell (as kinetic energy) located at the boundary between the BH (innermost $3M_\odot$ of the star) and the stellar envelope \citep{2017ApJ...846L..15B,2018ApJ...862L...3S}. These choices are informed by the spherically symmetric, general relativistic, hydrodynamical simulations of the same progenitor model by \citet{2012ApJ...754...76D}.

We model the companion BH as $15 M_\odot$ sink particle placed at an initial circular orbit with a separation $a=3R_\ast$ \citep{2018ApJ...862L...3S} and use a nearly isothermal equation of state ($\gamma= 1.1$) in order to capture the efficient cooling of the fallback material at these high accretion rates \citep{1993ApJ...411L..33C}. Strictly speaking, the inferred accretion rate, over the duration of the simulation (solid lines in Figure~\ref{fig:SN_time_evo}), remains above the neutrino-cooling mass accretion rate of $\dot{M}_{\rm hyper}\approx 10^4\dot{M}_{\rm Edd}\approx 10^{-4}M_\odot/{\rm yr}$ \citep{1991ApJ...376..234H}. As long as $\dot{M}\gtrsim \dot{M}_{\rm hyper}$, the inner regions of disks with mass fluxes above this limit are generally able to cool by neutrinos on time scales shorter than the inflow time \citep{1996ApJ...459..322C}.

Figure~\ref{fig:SN_time_evo} depicts the events preceding accretion disk formation. Initially, the supernova explosion is able to eject most of the outer layers of the progenitor star but fails to unbind its inner layers \citep{2017ApJ...846L..15B,2018MNRAS.476.2366F}. The corresponding fallback material is torqued by the binary and subsequently settles into a rotationally supported structure around each of the orbiting BHs. 

\begin{figure}
    \centering
    \includegraphics[width=1.0\linewidth]{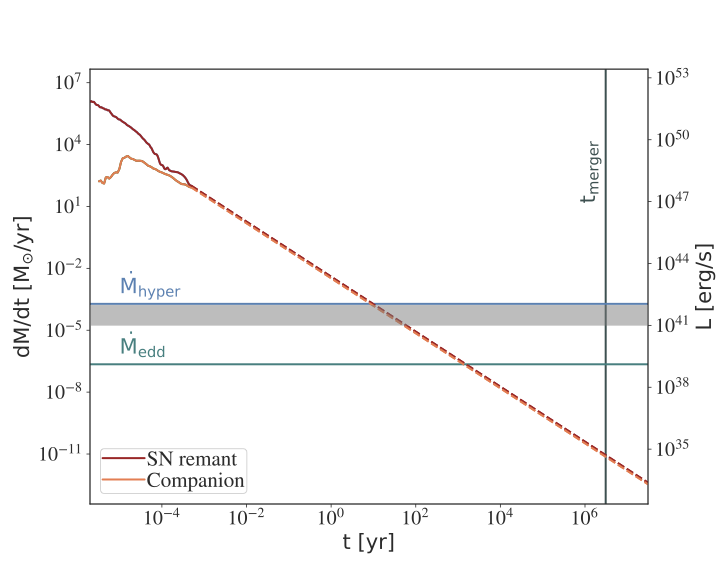}
    \caption{The accretion rate from the supernova fallback simulations presented in Figure~\ref{fig:SN_time_evo}  (solid lines) and the long-term accretion based on a power law decay extrapolation (dashed lines): $\dot{M} \propto t^{-\beta}$, with $\beta \approx 1$. The gray vertical line shows the time of merger expected from the newly assembled BH binary, while the horizontal blue and green lines show the lower limit for hypercritical neutrino-cooled accretion \citep{1993ApJ...411L..33C}  and the photon Eddington accretion limit, respectively. The right-hand-side $y$-axis show the accretion luminosity $L=\epsilon \dot{M}c^2$ with $\epsilon=0.1$. The gray-shaded region represents the peak luminosity range of the kilonova associated with GW170817 and short $\gamma$-ray bursts \citep{2017Sci...358.1583K,2019MNRAS.486..672A}
. }
    \label{fig:accretion_fallback}
\end{figure}

Figure \ref{fig:accretion_fallback} shows the mass accretion rate onto the individual BHs. Initially, the bound inner layers have very little angular momentum, and they are accreted directly onto the newly formed BH (denoted as {\it SN remnant}). The interaction of the fallback gas with the binary transfers orbital angular momentum to the infalling gas, thus subsequently halting the quasi-spherical accretion onto the newly born BH. This torqued material remains bound to the binary system and forms disks around both BHs (which are designated as {\it SN remnant} and {\it Companion} in Figure \ref{fig:accretion_fallback} and in t). As material with higher angular momentum settles around each of the BHs, the rate slows down and assumes a power law decay  (with index $\beta \approx -1$), as estimated for impulsively formed, viscously drained accretion disks \citep{1990ApJ...351...38C,2007NJPh....9...17L,2008MNRAS.390..781M}.  Obviously, the above power-law extrapolation is only a rough approximation and should be taken as an order of magnitude estimate for the mass of the disk at the time of merger. This is because drastic changes in the microphysics of the evolving disks are expected to occur as they transition from a neutrino-cooling-dominated regime to a photon-dominated regime.

At the end of the hydrodynamical simulation, the masses of the disks are $0.06M_\odot$  and  $0.13M_\odot$ around the SN remnant and the companion, respectively. The black hole binary has a post-supernova separation of $a_0 \approx 3.2 \ R_\odot$. The  merger time is generally given by \citep{peters1964mergertime} 
\begin{equation} \label{eq:t_p}
    t_{\rm merger} = \frac{5 a^4 c^5}{256 G^3 M_1 M_2 \left( M_1 + M_2\right)},
\end{equation}
where $a$ is the binary separation, and $M_1$ and $M_2$ are the masses of the companion and SN remnant BHs in the binary system, respectively. 
These are $M_{1}=15M_{\odot}$, $M_{2}=12.3M_{\odot}$, $a=a_0\approx3.2R_{\odot}$ at the end of the binary simulation. We estimate a merger time for the BH binary depicted in Figure~\ref{fig:SN_time_evo} to be $t_{\rm merger} (a= a_0)\approx$ 3.16 Myr (solid black vertical line in Figure~\ref{fig:accretion_fallback}). Using the self-similar solution depicted in Figure~\ref{fig:accretion_fallback}, we expect the BHs to have a luminosity of about $L_{\rm acc} (t= t_{\rm merger}) \approx 10^{35} $ erg/s around the merger time, under the assumption that the transport processes stay unchanged.

A detectable luminosity at the time of merger requires that the surrounding disk should not completely drain into the BH before the merger. Whether a disk of given mass remains depends sensitively not only on internal viscous dissipation and cooling (which likely alters the extrapolation made in Figure~\ref{fig:accretion_fallback}) but also on the effect of tidal forces (Section~\ref{sec:result}).
The hydrodynamical calculations presented in this section are used not only to provide an estimate of the steady-state luminosity at the time of merger but also serve to construct informed initial conditions for the Athena++ simulations presented in Section~\ref{sec:numerical}. The reader is referred to Section~\ref{sec:overview} for an overview of the evolution of the binary from the formation of the accretion disk around $M_1$  to the orbiting debris being truncated by the BH binary companion: $M_2$. The  effects of tidal forces on the relic disk around $M_1$ is expected to occur when the binary reaches a separation of about $a=a_\tau \approx 2.24R_\odot$. At $a=a_\tau$, tidal forces quickly truncate the accretion disk to radii of order half of the Roche lobe radius \citep{blondin2000tidally} and excite a two-armed spiral wave that remains stationary within the rotating reference frame of the binary system. But before turning our attention to exploring the role of spiral wave excitation at wider separations, in the following section, we provide a simple analytical estimate of how the steady-state luminosity (Figure~\ref{fig:accretion_fallback}) could be significantly enhanced at the final stages of the merger. This estimation assumes that  the surrounding disk was not otherwise dispersed before the merger and survived the enhanced angular momentum transport phase driven by tidal excitation, which is discussed in Sections \ref{sec:overview} and \ref{sec:result}. 

\subsection{Flaring counterparts to binary BH mergers}\label{sec:acrit}
The exact luminosity of the accompanying electromagnetic counterpart will depend sensitively on the accretion history of the disks around the BHs \citep{2016ApJ...821L..18P}, and on how much disk mass is left after a time $t_{\rm merger} (a=a_0)$. As the accretion disks deplete, the orbit of the binary secularly decays via gravitational wave radiation. There will come a point at which the standard viscous accretion time around the BH  will be equal to the gravitational wave inspiral time. If we estimate the viscous accretion timescale as 
\begin{equation} \label{eq:t_v}
    t_\nu = 2 \pi \sqrt{\frac{R^3}{GM_1}} \alpha^{-1} \left( \frac{H}{R}\right)^{-2}
\end{equation}
where $\alpha$ represents the standard viscosity parameter, $H/R$ is the scale height of the disk and $t_{\rm orb} =2 \pi \sqrt{R^3/(GM_1)}$ is the corresponding orbital period, we can then equate (\ref{eq:t_p}) to  (\ref{eq:t_v}) by setting $R\approx a$. This allows us to derive  an expression for the critical binary separation, $a_\text{crit}$, at which $t_{\rm merger}\approx t_\nu$:
\begin{equation}
   a_\text{crit} = \left[ \frac{512 G^{5/2} M_1^{1/2} M_2 \left( M_1 + M_2 \right) }{5 \pi c^5 \alpha \left( H/R \right)^{2} } \right]^{2/5}.
   \label{eq:acrit}
\end{equation}
A simplified form of expression (\ref{eq:acrit}) can be found when $M_1=M_2=M$:
\begin{equation*}
    a_\text{crit} \approx 264 R_\text{Sch}\left( \frac{\alpha}{0.01} \right)^{-2/5} \left( \frac{H/R}{0.1} \right)^{-4/5},
\end{equation*}
where $R_{\rm Sch}$ is the Schwarzschild radius of the BH. Here $a_\text{crit}\ll a_\tau <a_0$. At this critical separation $t_{\rm merger}\approx t_\nu$, where
\begin{equation*}
 t_\nu (a_\text{crit}) \approx 5.6 \times 10^4  \left( \frac{M}{15 \ M_\odot}\right) \left( \frac{H/R}{0.1} \right)^{2} \left( \frac{\alpha}{0.01} \right)^{-1}\;{\rm s},
\end{equation*}
and 
\begin{equation*}
 t_{\rm orb} (a_\text{crit}) \approx 5.6  \left( \frac{M}{15 \ M_\odot}\right)\;{\rm s}.
\end{equation*}

The disk mass contained within $a_\text{crit}<R_{\rm disk}$ is expected to be drained dynamically soon after the merger of the binary \citep{2016ApJ...821L..18P,2017ApJ...839L...7D}. This is mainly because the burst of gravitational wave emission during the last stages of the merger results in a corresponding, nearly instantaneous reduction in the binary's rest mass, which, in turn,  excites a single-armed spiral shock wave that can transport angular momentum effectively \citep[e.g.,][]{2010MNRAS.404..947C}. 
We solve for the disk mass by integrating the disk density profile \citep{2002apa..book.....F}, assuming an accretion rate of $10^{-11} \ M_\odot / \rm yr$ as predicted by Figure \ref{fig:accretion_fallback} at $t=3.16 $ Myr. The total disk mass can be written as 
\begin{equation}
    M_{\text{disk}} = \int_0^{a_{\text{crit}}} 2 \pi R \Sigma dR ,
\end{equation}
where 
\begin{equation}
    \Sigma = 5.2 \alpha^{-4/5} \dot M_{16}^{7/10} M_1^{1/4} R_{10}^{-3/4} 
    \left[ 1 - \sqrt{ \frac{R_{\text{BH}}}{R}} \right]^{7/10}.
\end{equation}
Here $\dot M_{16} = \dot M / (10^{16} \ \rm g \ s^{-1})$, $R_{10} = R / (10^{10} \ \rm cm)$ and we have assumed that the disk is gas pressure dominated and that the Rosseland mean opacity is well approximated by Kramers’ law ($T\lesssim 10^6$K and $\rho\lesssim 10^{-10}$ g/cm$^3$). This yields a total disk mass of $\approx 10^{-12} M_\odot$ given the accretion disk conditions expected at $t=3.16 $ Myr. In principle, the flow onto the BH  can liberate gravitational potential energy at a rate approaching $\epsilon \dot{M} c^2$. An optimistic estimation of the flow inflow rate, $\dot{M}$, can be derived by conjecturing that the disk mass within $a_{\rm crit}$ drains dynamically. In this case $\dot{M} \approx M_{\rm disk} (r< a_{\rm crit})/t_{\rm orb}(a_{\rm crit})$, which  yields a luminosity with an efficiency $\epsilon=0.1$  that can be written as     
\begin{equation}
   L_{\rm flare} \approx 3 \times 10^{40} \left(\frac{M_{\rm disk}}{10^{-12} \ M_\odot}\right) \left( \frac{M}{15 \ M_\odot}\right)^{-1}\;{\rm erg/s}.
\end{equation}
For completeness, we can also write the associated temperature of the flare as the black body temperature if a luminosity $L_{\rm flare}$ emerges from a disk surface of radius $R_{\rm Sch}$: $T_{\rm flare} \approx 3 \times 10^7$K. While the flaring accretion luminosity is expected to be significantly larger than the pre-flare, steady-state luminosity depicted in Figure~\ref{fig:accretion_fallback},
this brighter signal is not only orders of magnitude fainter than the kilonova (Figure~\ref{fig:accretion_fallback}) associated with GW170817 \citep{2017Sci...358.1583K} but is also expected to peak at much higher photon energies, where it becomes harder to detect fainter sources due to factors like reduced detector efficiency and increased background noise.  These attributes, combined with the short duration of the signal, make the detection of the flare prohibitively difficult. 

\begin{figure*}   \includegraphics[width=\textwidth]{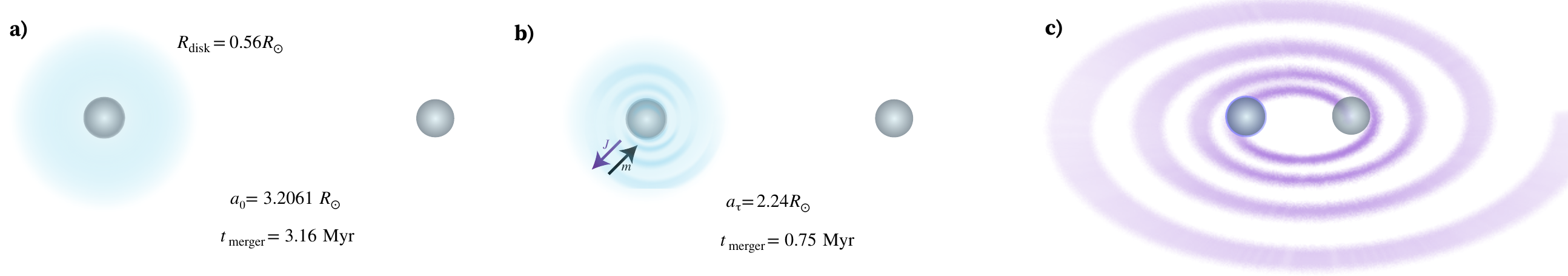}
    \caption{Diagram illustrating the evolution of a relic accretion disk in a merging BH binary.  Disk formation is assumed to result from the formation of the second BH in the binary in a fallback SN. The simulation results presented in Figure~\ref{fig:accretion_fallback} are used to inform the conditions of the gas in and around the binary (panel {\it a}). At this stage, $t_{\rm merger} (a=a_0)=3.16$ Myr.  Here we focus on the evolution of the more massive disk, which is found to be around the companion. The evolution of the disk is expected to be altered when the binary has reached a separation at which tidal forces on the disk will impact angular momentum transport (panel {\it b}). This takes place as the binary shrinks via gravitational wave emission and $a=a_\tau\approx 2.24R_\odot$. At this separation, the accretion disk, whose initial circularization radius $R_{\rm disk}\approx 0.56 R_\odot$, will be effectively truncated due to the tidal force from the $M_2$ BH. At this point, $t_{\rm merger} (a=a_\tau)=0.75$Myr.  By the time the merger takes place, we expect the orbiting BHs to merge in a gas-free environment (panel {\it c}). However, if a relic disk is able to endure, the binary will evolve until $a_{\rm crit} \ll a_\tau$. At this final stage, the viscous accretion time around the BH is expected to be similar  to the gravitational wave inspiral time, $t_{\rm merger} (a=a_{\rm crit})\approx 15.5 $ hours, potentially resulting in an  electromagnetic flaring signature (Section~\ref{sec:acrit}).
    }
    \label{fig:diagram}
\end{figure*}
It is thus evident from the above discussion that the eventual resulting electromagnetic signature depends fairly strongly on the properties of the disk at the time of merger and, as argued by many authors, its detection will benefit from the field binary hosting a massive relic disk at the time of merger. This has stimulated the community to try to explore physically-motivated ways of decreasing the rate at which angular momentum is transferred outwards within the disk, effectively slowing down the process of matter accreting onto the BH \citep{2016ApJ...821L..18P}. However, all currently discussed scenarios in the literature neglect the potentially critical influence of the BH companion, for which tidal forces on the disk will impact disk draining 
\citep{paczynski1977model, 1977MNRAS.181..441P}. It is to this issue that we now turn our attention. In what follows we  study, for the first time,  the effect of hydrodynamic shocks on angular momentum transport in the vicinity of the companion, which can potentially enhance the accretion rate in the relic disk and, therefore, reduce the pre-merger disk mass.

\section{The Evolution of an Accretion Disk in a Merging Black Hole Binary}\label{sec:overview}

\subsection{Setting the stage: From disk assembly to merger}
A schematic montage of successive interactions, which are believed to be central to the assembly and evolutionary stages of a relic disk in a merging BH binary, is shown in  Figure~\ref{fig:diagram}. The different frames show the inferred orbital arrangement of the binary as a function of time. Let us consider these frames in turn, working from the formation of the accretion disk, which occurs after the secondary massive star explodes (Figure~\ref{fig:SN_time_evo}), to disk draining, which is induced by the torque from the BH companion as the binary contracts and the accretion disk orbiting the primary BH is truncated (Figure~\ref{fig:spiralplots}). This is expected to take place at $a\gg a_{\rm crit}$ (Section~\ref{sec:acrit}).

The binary scenario envisioned in this study and depicted in Figure~\ref{fig:diagram} commences with a tight binary ($a=a_0\approx3.2R_\sun$) composed of a $15 M_\odot$ BH orbiting a 12.3$M_{\odot}$ BH. For the sake of simplicity, we follow the evolution of the more massive accretion disk, which is assembled by the SN fallback of the secondary star around the primary  $15 M_\odot$ BH (panel {\it a} in  Figure~\ref{fig:diagram}). This is because the disk around $M_1$, the more massive BH, is the last one to be tidally excited before the merger and, as such, the orbiting mass that is expected to potentially remain before the merger. The disk surrounding $M_2$, the lower mass BH, will experience tidal excitation at an earlier evolutionary stage. Yet, we expect the results of our numerical experiments to be widely applicable to understanding the evolution of any relic disk when a BH binary reaches a separation at which tidal forces on the orbiting disk become prominent. 

The initial mass and size of the accretion disk are $M_{\rm disk}=0.13M_\odot$ and $R_{\rm disk}=0.56R_\odot$, respectively (Figure~\ref{fig:SN_time_evo}). The merger time at this stage is $t_{\rm merger}(a=a_0)=3.16$ Myr, while the viscous accretion timescale at the outer disk radius is $t_\nu \ll t_{\rm merger}(a=a_0) $. As a result, the disk is expected to be highly depleted by the time the binary merges via gravitational wave emission. However, many authors advocate for the existence of a massive remnant disk, given that the evolution of these remnant disks is highly uncertain \citep{2016ApJ...821L..18P,2019ApJ...875...49P, 2017ApJ...839L...7D,2016ApJ...822L...9M,2016ApJ...823L..29K,2018MNRAS.480.4732M}. The main argument given in the literature in support of this is that the disk might cool below the temperature at which electrons are ionized \citep{2016ApJ...821L..18P,2019ApJ...875...49P} above which the MRI is thought to operate effectively. 

In the next sections we show that even in the case of inefficient MRI transport, the angular momentum transport in the last stages is likely to be dominated by spiral shocks. This occurs when the binary reaches a separation of about $a=a_\tau \approx 2.24R_\odot$ (which in the code units used in Section~\ref{sec:ic_bc} corresponds to $a_\tau=1$). At this separation, the accretion disk, whose initial circularization radius is $\approx 0.56R_\odot$, is effectively truncated due to the tidal force from the BH companion ($M_2=12.38M_\odot$). The torque from the companion perturbs 
the disk and excites stationary spiral shocks, which in turn drains the disk, even in the absence of other angular momentum transport mechanisms. The merger time at the disk-draining stage is $t_{\rm merger} (a=a_\tau) =0.75$ Myr. The gas-depleted binary is then expected to merge via gravitational wave emission (panel {\it c} in  Figure~\ref{fig:diagram}). We note that this draining stage occurs well before the separation at which the disk is unable to viscously react to the gravitational wave orbital hardening (Section~\ref{sec:acrit}). That is $a_\tau \gtrsim a_{\rm crit}$, where $t_{\rm merger} (a=a_{\rm crit}) =15.5$ hours $\ll$ $t_{\rm merger} (a=a_\tau) =0.75$ Myr. 

The goal of the following numerical experiments is to gain a deeper physical interpretation of the evolution of relic accretion disks in BH binaries.  Because a single simulation of the full problem incorporating all of the aforementioned effects would  be prohibitively expensive, the understanding of the evolution of accretion disks around BH binaries requires a novel approach that attempts to resolve the underlying physics on a wide range of scales. Motivated by this, we propose to study the evolution of tidal excited accretion disks  via a series of numerical experiments that isolate the key processes that regulate the transport of angular momentum via spiral waves. In addition to being computationally feasible, this approach will enable a thorough understanding of the relevant processes. It is important to note that in this study we neglect the viscous spreading of the disk from assembly to the separation at which tidal truncation occurs. This is  a justifiable assumption given that the vast majority of the mass remains near the initial circularization radius and only a small amount of mass is able to spread viscously beyond  $R_{\rm disk}$ \citep{1990ApJ...351...38C, 2015ApJ...804...87L}. 

\subsection{Methods and Numerical Framework}
\label{sec:numerical}
We present two-dimensional, viscous hydrodynamic simulations using Athena++ \citep{stone2020athena++} in cylindrical coordinates. Hereafter, we use $R$ for the radial direction coordinate and $\phi$ for the polar angle direction coordinate. Our numerical framework includes two stages. First, we simulate the mass transfer from the SN remnant in order to form a disk around the companion BH, which we refer to as the disk-building phase. To do this, we adopt the widely used setup for studying accretion disks in mass transferring close binaries \citep[e.g.,][]{blondin2000tidally,makita2000two}. We model the Roche lobe overflow as a mass stream injected through the L1 point at the outer boundary. We then evolve the disk until a quasi-steady state is established. In the next stage, we turn off the mass feeding from the L1 point at the boundary, which allows us to study the accretion of the post-supernova relic disk and the angular momentum transport. In the rest of the \textit{Letter}, we refer to this second stage as the disk-draining stage.  

We found that the column density radial profile of the  constructed orbiting debris in the disk-draining stage closely resembles the conditions found in the post-supernova relic disk, where $\Sigma \propto r^{-\alpha}$ and $\alpha \approx -3/10$ (Figure~\ref{fig:SN_time_evo}). Similar radial profiles are also found in the solutions used to describe the structure and evolution of impulsively assembled disks \citep{1990ApJ...351...38C, 2015ApJ...804...87L}. As such, the formalism described in Section~\ref{sec:ic_bc} can be used to  effectively construct the initial conditions of impulsively assembled accretion disks, which are necessarily   radially truncated.

\subsection{Equations and Scalings}\label{sec:eq_scaling}
The conservation equations solved by Athena++ are as follows: 
\begin{eqnarray}
 \partial _t \rho + \nabla \cdot( \rho \mathbf{v}) &  = & 0   \nonumber \\
 \partial _t (\rho \mathbf{v}) +   \nabla \cdot (\rho\mathbf{vv}+\mathsf{P}-\mathsf{\Pi})   & = &-  \rho \mathbf{a}_{\rm{ext}}  \nonumber \\
 \partial _t E +   \nabla \cdot( [ E + P] \mathbf{v}) & = & - \rho \mathbf{a}_{\rm{ext}} \cdot \mathbf{v},
\label{eq:fluid_eq}
\end{eqnarray}
where $\rho$, $\mathbf{v}$, $E=E_{\rm g}+(1/2)\rho v^{2}$ are fluid density, velocity and total energy density. In the adiabatic simulations, the gas internal energy is related to gas pressure by $E_{\rm g}=P/(\gamma-1)$, where $\gamma$ is the gas adiabatic index. 

In the momentum equation, $\textsf{P}$ is the pressure tensor. For our 2D hydrodynamical simulations, we include contributions from the viscous stress tensor $\mathsf{\Pi}$ \citep{2020ApJS..249....4S} to roughly approximate the angular momentum transport through turbulence driven by the magnetorotational instability \citep[hereafter MRI]{1991ApJ...376..214B}.
\begin{equation}
    \Pi_{ij}=\rho\nu\left(\frac{\partial v_{i}}{\partial x_{j}}+\frac{\partial v_{j}}{\partial x_{i}}-\frac{2}{3}\delta_{ij}\nabla\cdot\textbf{v}\right)
\end{equation}
where $\nu$ is the kinematic viscosity coefficient. The standard $\alpha$-disk description viscosity $\alpha$ is related to the kinematic viscosity coefficient by $\nu=\alpha c_{\rm s}H$.
On the right-hand side, $\rho\mathbf{a}_{\rm ext}$ are external forces including gravity and non-inertial forces:
\begin{eqnarray}
\mathbf{a}_{\rm{ext}} &= - \frac{GM_{1}}{|  \mathbf{R}|^{3}}  \mathbf{R} - \frac{GM_{2}}{|  \mathbf{R} -  \mathbf{R_{2}} |^{3}} ( \mathbf{R} -  \mathbf{R_{2}}) + \mathbf{a}_{\rm frame},\\
&\approx  \frac{GM_{1}}{|  \mathbf{R}|^{3}}  \mathbf{R} - f_{\rm soft} (M_{2}, \mathbf{R} -  \mathbf{R_{2}}) + \mathbf{a}_{\rm frame},
\label{eq:sourceterms}
\end{eqnarray}

Our simulations are centered on the companion BH, with the SN remnant BH outside of the computational domain. We assume that the binary has reached a separation at which tidal forces on the disk will impact accretion and disk draining. This is expected to occur at a binary separation of $a_\tau\approx 2.24 \ R_\odot$, given the post-supernova disk radius $\approx 0.56 \ R_\odot$ (Figure~\ref{fig:SN_time_evo}). The merger time at this moment in the life of the binary is $t_{\rm merger} (a=a_\tau) \approx 0.75$ Myr while the viscous accretion timescale at the outer disk radius is $t_\nu \approx 0.5$ years, assuming $\alpha=0.01$ and $(H/R)=0.1$ (Equation~\ref{eq:t_v}). Significantly longer viscous accretion timescales have, however, been envisioned for cold relic disks \citep{2016ApJ...821L..18P}, in which the disk is conjectured to cool below the temperature at which electrons are ionized  and the viscous transport is thought to operate inefficiently. 

We approximate the gravity of $M_{2}$ as a softened point mass $f_{\rm soft} (M_{2}, \mathbf{R} -  \mathbf{R_{2}})$ with softening kernel $f_{\rm{spline}} $ from \cite{1989ApJS...70..419H} and adopt a softening radius $r_{\rm{soft}}=0.1$, so the gravity of $M_{2}$ in the simulation domain of this work is in the Newtonian regime. Although it is not strictly necessary to include a softening potential for $M_{2}$, as it resides outside of the simulation domain, we include it for consistency with follow-up works that will include the companion within the domain.

We perform simulations in the frame centered on $M_{1}$ and co-rotating with both objects at an angular speed $\Omega=\sqrt{(GM_{1}+GM_{2})/R_{2}^{3}}$. Accounting for these frame transformations, the non-inertial forces are:
\begin{equation}
\mathbf{a}_{\rm frame} =  - \Omega \times (\Omega \times \mathbf{R}) -  2 \Omega \times \mathbf{v} + \frac{GM_{2}}{|\mathbf{R_{2}}|^{3}}\mathbf{R_{2}}.
\label{eq:non-inertial}
\end{equation}
The first two terms in Equation~\ref{eq:non-inertial} are the centrifugal and Coriolis forces, respectively. The third term is from the transformation between the center-of-mass frame and the $M_{1}$-centered frame \citep{binney2011galactic}.

We solve dimensionless equations by adopting the following units of mass $m_{0}$, length $l_{0}$, and time $t_{0}$. The unit of mass is the total mass of the system $Gm_{0}=GM_{1}+GM_{2}=1.0$, with mass ratio $q \equiv M_{2}/M_{1}$. The unit of length is set to the separation of the binary $L_{0}=a$, so the companion is at $\mathbf{R_{2}}=\mathbf{\hat{x}}$. The unit of time is related to the angular frequencies by $t_{0}=1/\Omega$, making one orbital period $P_{\rm 0}/t_{0}=2\pi$. For fiducial parameters, we set $M_{1}=15M_{\odot}$, $M_{2}=12.3M_{\odot}$, $a_\tau=2.24R_{\odot}$ regarding to the end of the binary simulation (i.e., $l_0=a_\tau=1$ in code units). With this set of units, we have the ability to rescale the simulations to various binary systems. In the rest of the paper, we report results in dimensionless units using the above scalings unless otherwise specified.

\subsection{Initial and boundary conditions}\label{sec:ic_bc}
We model the mass transfer from the companion as gas injected through the L1 point at the simulation domain boundary. The calculation domain spans as $(R_{\rm min}, R_{\rm L1})\times(0, 2\pi)$ in the R- and $\phi$-directions, where $R_{\rm min}=0.008$ is the inner radius, $R_{\rm L1}\approx 0.520$ is approximately the distance between the $M_{1}$ and L1 point for our fiducial $q=0.82$. For our given scaling, where the length unit is set by the separation of the binary, the inner boundary location of $R_{\rm min}=0.008$ is  well outside of the inner stable circular orbit (ISCO) for the BH. In the rest of the paper, the quantities measured near the inner boundary do not necessarily reflect the physics near ISCO. We adopt a logarithmic grid with 384 cells in the R-direction and a uniform grid with 704 cells in the $\phi$-direction to maintain the cell aspect ratio at larger radii. Initially, we set the uniform density and pressure in the domain $\rho_{\rm init}=\rho_{\rm floor} = 10^{-4}$ and  $p_{\rm init}=p_{\rm floor} = 10^{-6}$.  We adopt a locally isothermal temperature profile throughout the disk, such that the gas temperature at each radius $T_{\rm gas}(R)=(GM_{1}/R)(H/R)^{2}$. With this temperature profile and for Keplerian velocity, the kinematic viscosity coefficient $\nu=\alphak T_{\rm gas}\sqrt{R^{3}/GM_{1}}$. In simulations, the equation of state is adiabatic with $\gamma=5/3$, the temperature profile is inserted as a temperature upper limit, in practice, this yields roughly constant Mach number $\mathcal{M} \approx(H/R)^{-1}$ in the disk, which provides a channel to compare with theoretical dispersion relation.

We adopt periodic boundaries in the $\phi$-direction (azimuthal). For the inner $r$-direction boundary, we copy all the variables from the last active zone but set any positive radial velocity to zero in the ghost zones. We treat the outer R-direction boundary condition differently for the two stages. In the disk-building stage (the first stage), if $\phi<0.1$ or $2\pi-\phi<0.1$, we treat the cell as L1 point and set the ghost zone density $\rho_{\rm gh}=1.0$, radial velocity $v_{\rm r,gh}=-0.01$, azimuthal velocity $v_{\rm \phi,gh}=0.0$ and pressure $p_{\rm gh}=0.01$ to model the mass feeding from companion. The mass-feeding stream parameters are chosen to form a general circularizing, tidally truncated disk in a binary system. We mention in Section~\ref{sec:discussion} that the angular momentum transport is not sensitive to the exact mass of disk with this set-up. For other cells, we use a single-direction outflow boundary by copying the variables from the last active zone and setting any negative radial velocities to zero. In the disk-draining stage, we adopt a single-direction outflow boundary for all cells at the outer $r$-direction boundary, so there is no mass feeding to the system. 

\begin{figure*}
    \centering
    \includegraphics[width=0.9\textwidth]{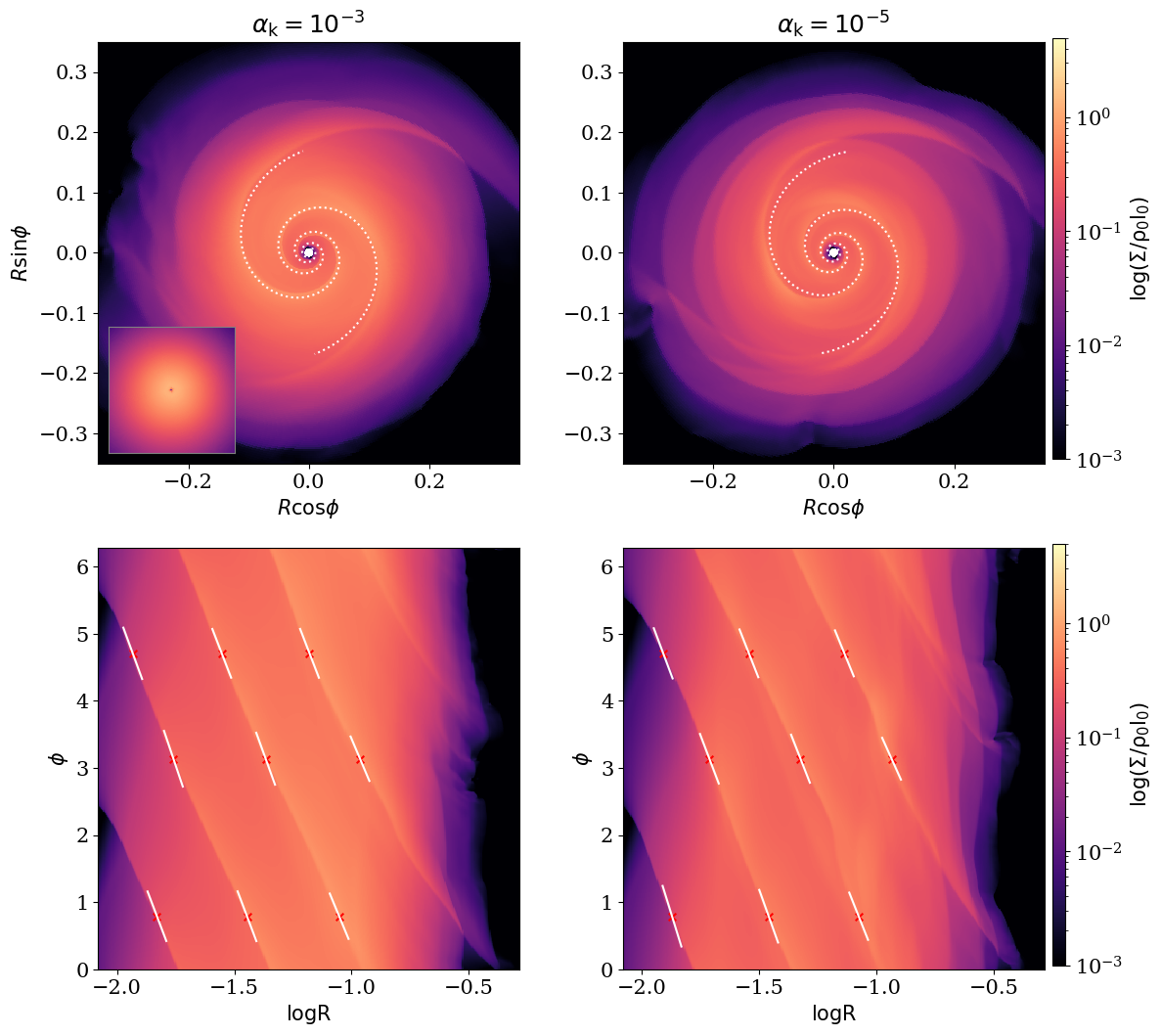}
    \caption{Spiral shocks in binary black holes. {\it Top} panel: gas density snapshots in the orbital plane at $t=200$ for  \textsf{Am3} ({\it left}) and \textsf{Am5} ({\it right}). The inset panel in \textsf{Am3} shows the gas surface density of \textsf{Am3\_nc} at the same time, where we set the companion BH's gravity $GM_{2}=0.0$. {Bottom} panel: gas surface density snapshots of \textsf{Am3} ({\it left}) and \textsf{Am5} ({\it right}) plotted in the $\log R-\phi$ coordinate system to illustrate the pitch angle of the spiral shocks. We sample the Mach number, $\mathcal{M}$, and rotation speed, $\Omega(r)$, locally at the points marked by the red cross symbols. The white line segments are pitch angles fitted with linear wave theory as in Equation~\ref{eq:pitchangle}. The white dashed lines in the {\it top} panels show the spiral patterns fit using the average Mach number and rotation speed from the red cross symbols in the {\it bottom} panels.}
    \label{fig:spiralplots}
\end{figure*}

\section{Spiral Structures and Shocks in Accretion Discs in
Black Hole Binary Systems}\label{sec:result}
\subsection{Spiral Shock Formed in the Disk}
In this section, we report the results of four simulations that share the same disk building-up phase. We adopt $\gamma=5/3$ and $H/R=0.2$ \citep[which is typical for disks at low accretion rates;][]{2022iSci...25j3544L} for the build-up phase and keep these parameters unchanged in the disk-draining phase. We run the disk-building stage up to $t=40$, roughly corresponding to $\approx 13$ binary orbital periods. The disk-draining stage lasts for $t\approx2 00$ after the building stage, corresponding to $\approx 64$ binary orbital periods. We stop the disk-draining simulations when the disk mass drops by at least $60\%$ mass. For the purpose of studying angular momentum transport, it is common to evolve the disk until the spiral wave pattern is stationary instead of when the accretion rate reaches steady-state \citep{blondin2000tidally}. We set $\alphak=10^{-5}$ in the build-up stage, which is negligible compared to the contribution from spiral shocks we measured in the simulations. 

We measure the disk draining of the companion with and without the presence of the BH companion to test the effect that spiral shocks excited by the remnant have on the disk's evolution. Throughout this \textit{Letter}, simulation runs with a subscript of $\textsf{\_nc}$ refer to the `no companion' tests. We additionally vary $\alphak$ from $\alphak=10^{-5}$ to $\alphak=10^{-3}$ to test the role of viscous dissipation from spiral shocks on the disk draining process when compared to the standard disk viscosity. Simulation runs with  $\alphak=10^{-5}$ ($\alphak=10^{-3}$) are thus referred to as \textsf{Am5} (\textsf{Am3}) when the gravity of the BH companion is included and \textsf{Am5\_nc} (\textsf{Am3\_nc}) when $GM_{2}=0.0$. 

In the disk build-up phase, the feeding stream impacts the disk and effectively dissipates energy. Consistent with previous studies, the stream quickly circularizes and forms an accretion disk. The disk truncates around $R_{\rm disk}\approx 0.25a_\tau$ ($a_\tau=1$ in code units)  due to the tidal force from the companion (Figure~\ref{fig:spiralplots}), which is about 60\% of the Roche lobe radius for the binary system \citep{blondin2000tidally}. After $\approx 13$ binary orbital periods, when the average angular speed of gas in the disk is close to the Keplerian value, the spiral wave pattern is observed to be stationary. At this point, we turn off the stream and initiate the disk-draining phase. 

Figure~\ref{fig:spiralplots} shows snapshots of the gas density in the orbital plane in the disk-draining phase. In simulation \textsf{Am3} and \textsf{Am5}, the torque from the companion perturbs the disk and excites stationary spiral shocks. In contrast, the spiral shocks disappear in the absence of a companion. This is clear in \textsf{Am3\_nc}, which is shown in the inset panel of Figure~\ref{fig:spiralplots}, whose domain spans the same range as the larger figure. The density distribution of this isolated disk is smooth, and the velocity profile is Keplerian without tides. Without tidal truncation, the isolated disk expands to a larger size compared to the disk with a companion due to the inserted viscosity $\alpha_{K}$. The isolated disks show higher surface densities than their counterparts when the BH companion is included (\textsf{Am3} and \textsf{Am5}).

The spiral waves will approximately follow the linear dispersion relation if shock dissipation is not too large \citep[see e.g.][]{1987gady.book.....B, ogilvie2002wake} as in our simulations. We show in Figure~\ref{fig:spiralplots} that the spiral shocks seen in \textsf{Am3} and \textsf{Am5} closely follow the linear wave dispersion relatio, the reader is referred to Appendix~\ref{appendix:spiralshock} for the corresponding description and further discussion. The pitch angle of such spiral shocks in a two-dimensional, Keplerian-like disk is given by Equation~\ref{eq:pitchangle} in Appendix~\ref{appendix:spiralshock}, where the local Mach number is the leading factor. The white line segments plotted in the $\log R-\phi$ coordinate system in Figure~\ref{fig:spiralplots} illustrate the pitch angle according to Equation~\ref{eq:pitchangle} with $m=2$ for the azimuthal wavenumber mode. The Mach number $\mathcal{M}$, and rotation speed, $\Omega(r)$, are locally sampled at various locations, which are marked by the red crosses in the right columns Figure~\ref{fig:spiralplots}. We also plot the pitch angles as white dotted line segments in the {\it top} panels in Figure~\ref{fig:spiralplots}. As argued by \citet{ju2016global}, the dispersion relation derived by \citet{ogilvie2002wake} provides a first-order description for the spiral patterns emanating in tidally truncated accretion disks. 

For a Keplerian disk with a temperature profile of $T(R)=v_{\rm K}^{2}(H/R)^{2}$, $\mathcal{M}$ is independent of the radius and is proportional to the inverse of scale height $\mathcal{M} \approx(H/R)^{-1}$. In the inner radius ($R \lesssim 0.14$), the disk temperature closely follows $T(R)\propto(H/R)^{2}$, implying a roughly constant Mach number. In this region, we find that Equation~\ref{eq:pitchangle} approximates the spiral pitch angles relatively well. At the outer radii, the local velocity deviates from the Keplerian velocity due to the companion's torque. As a result, the dispersion relation given by Equation~\ref{eq:pitchangle} no longer effectively describes the pitch angle of the spiral shock. Therefore, we only show the fitted spiral patterns for the inner region in Figure~\ref{fig:spiralplots}. 

\subsection{Angular Momentum Transport by Spiral Shocks}\label{sec:am}
\begin{figure*}[htb]
    \centering
    \includegraphics[width=0.8\textwidth]{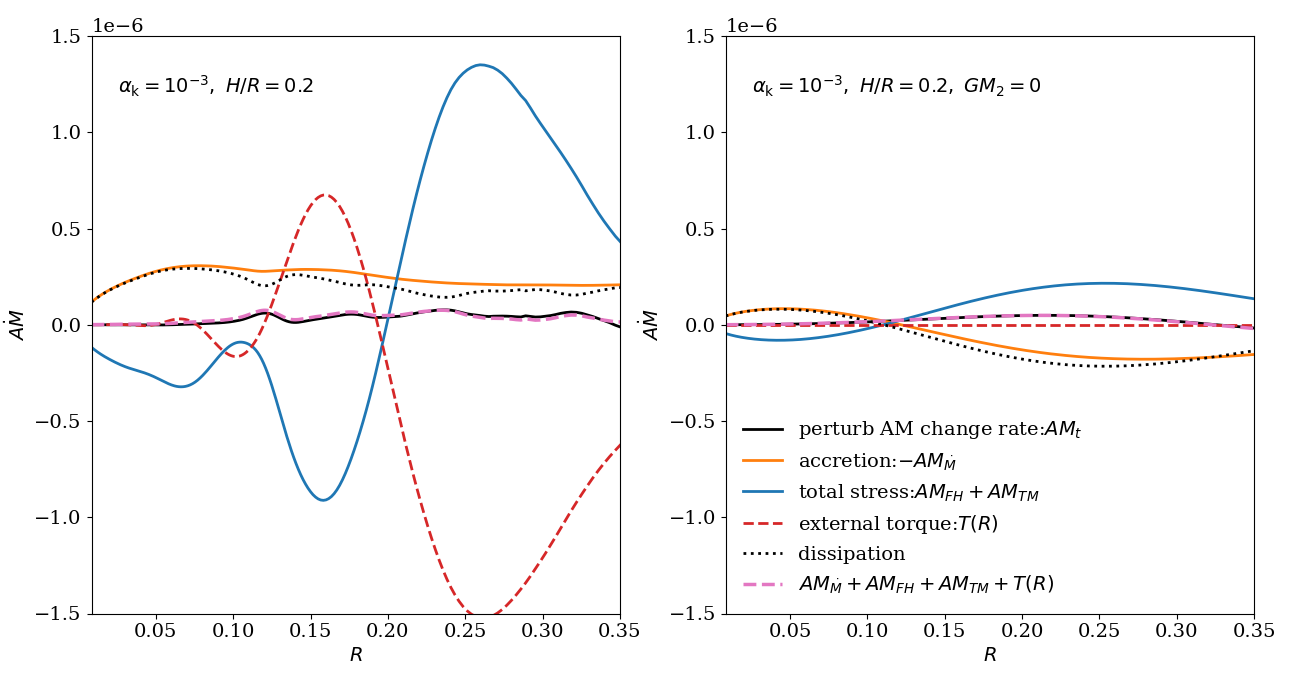}
    \caption{The perturbed angular momentum at a given radial distance in the non-corotating frame for a disk with (\textsf{Am3}, left) and without (\textsf{Am3\_nc}, right) a companion BH are shown in the {\it left} and {\it right} panel, respectively.  Components of the perturbed angular momentum terms are listed in the {\it right} panel and are derived in Appendix~\ref{appendix:amtransfer}. The results are time-averaged over the disk-draining phase. The black solid curves correspond to the rate of perturbed angular momentum change (Equation~\ref{eq:amdt}). The orange solid curves are associated with the angular momentum transport of the accreting mass flux (Equation~\ref{eq:ammdot}). The blue solid curve represents the Reynolds and the viscous stress terms (Equation~\ref{eq:ammdot}), and the red dashed curve corresponds to the torque from the companion (Equation~\ref{eq:amtorque}). The dotted black curves are the difference between the total stress and the torque, which we defined as dissipation that induces angular momentum transport and accretion. In the case of low $\alphak=10^{-5}$, where the hydrodynamics are nearly inviscid, the companion's presence leads to the non-zero torque term, and shock dissipation gives rise to mass accretion.}
    \label{fig:torques}
\end{figure*}
Here, we concentrate on the effect of global density waves on the angular momentum transport of a centrifugally supported disk in our simulations. The resulting waves, as observed in Figure~\ref{fig:spiralplots}, have reasonably large amplitudes, which, as a result of steepening, evolve rapidly into shocks. Dissipation at the shocks provides the channels of momentum and energy exchange between the wave and the disk \citep{rafikov2016protoplanetary}. 

To quantify the ensuing transport, we make use of the conservation equation for perturbed angular momentum derived by \citet{ju2016global} for a rotational supported flow, which is presented for completeness in Appendix~\ref{appendix:amtransfer} (see Equations~\ref{eq:perturbedAMeq}-~\ref{eq:amtorque}). The conservation equation for perturbed angular momentum can be written as
\begin{equation}
\text{AM}_{t}=\text{AM}_{\dot{M}} +\text{AM}_{\rm FH}+\text{AM}_{\rm TM}+ T, 
\label{eq:per_ang}
\end{equation}
where $\text{AM}_{t}(R)$ is the time derivative of perturbed angular momentum within a ring located at a radial distance $R$. It is close to zero if the disk reaches a quasi-steady state. On the right-hand side of the equation we have the angular momentum change associated to the total mass accretion rate passing through this ring $\text{AM}_{\dot{M}}(R)$, to the Reynolds stress $\text{AM}_{\rm FH}(R)$, to the viscous stress $\text{AM}_{\rm TM}(R)$ and to the torque exerted by the companion $T(R)$. As conferred in Appendix~\ref{appendix:amtransfer}, the two terms ($\text{AM}_{\rm FH}$ and $\text{AM}_{\rm TM}$) are both included in the $\phi$-component of the radial momentum flux in Athena++ and, as such, are calculated together. We list the detailed form for each term in Appendix~\ref{appendix:amtransfer} Equation~\ref{eq:perturbedAMeq}-Equation~\ref{eq:amtorque}. Note that the above equation and Equation~\ref{eq:perturbedAMeq} differ from the perturbed angular momentum equation in \citet{ju2016global} by a factor of $1/R$.

In the linear wave propagation regime, the radial advection of angular momentum terms must cancel the external torque, leading to a zero mass accretion rate. That is, the torques serve to restore the spiral structures. If the waves steepen into shocks, dissipation (irreversible heating) happens when a fluid element passes the shockfront, which is not aligned with circular orbits. From the perspective of angular momentum conservation, the role of dissipation can be quantified by the difference between the external torque and the angular momentum flux advection. In this case, when the disk reaches steady state, dissipation is the source of accretion. 

In Figure~\ref{fig:torques}, we show the angular momentum budget for the simulation run \textsf{Am3} ({\it left} panel), which is, for completeness, compared with the simulation run \textsf{Am3\_nc} of a disk evolving without a BH companion ({\it right} panel). The black solid curve shows the variation of perturbed angular momentum (Equation~\ref{eq:amdt}). Because the disk is not in a strict quasi-steady state, the term is not exactly zero but is considerably smaller when compared to other terms. The {\it pink dashed} curve shows the sum of all the terms affecting angular momentum in the simulation (Equation~\ref{eq:ammdot}-Equation~\ref{eq:amtorque}), which, as expected, matches the {\it black solid} curve. There are small differences due to the conversion between the co-rotating and non-rotating frames as well as the mixed usage of cell-centered variables and cell-face variables (Appendix~\ref{appendix:amtransfer}).

The blue solid curve corresponds to the total stress, including Reynolds and viscous stresses (Equation~\ref{eq:amth}), and the red dashed line corresponds to the torque from the companion (Equation~\ref{eq:amtorque}). If the linear waves propagate without dissipation, the angular momentum injected by the torque will be advected radially by the waves, which, in principle, would lead to the exact cancellation between the total stress term $AM_{\rm FH}+AM_{\rm TM}$ and the torque term $T(R)$. The fact that there is a net residual clearly illustrates that spiral shock dissipation results in a net accretion of gas, which is shown by the {\it orange} curve (Equation~\ref{eq:ammdot}). This shows that accretion in the presence of a companion is largely driven by the shock dissipation, especially within the disk truncation radius ($R\lesssim 0.25$). In contrast, $T(R)=0$ in the absence of the companion (\textsf{Am3\_nc}). The overall angular momentum transport is largely suppressed without the excitation of spiral shocks by the companion, thus delaying disk draining. We find similar scenario in \textsf{Am5}, where the parameterized $AM_{\rm TM}$ corresponds to negligible $\alphak=10^{-5}$. We thus conclude that the companion's presence will likely enhance the accretion rate and deplete disk gas rather effectively, even in the case of a nearly inviscid disk \citep{2016ApJ...821L..18P}. 

\section{Discussion and Future Prospects}
\label{sec:discussion}
Observational astronomy is in an era of large-scale, systematic exploration, routinely revealing new fascinating transients. Despite this advancement, one of the still missing pieces in gravitational wave astronomy is the definite identification of an electromagnetic counterpart to a binary black hole merger \citep{2011CQGra..28i4021S,2016ApJ...819L..21L, 2019ApJ...875...49P,2023ApJ...942...99G}.
In the merger of two black holes, an electromagnetic signal is not expected to be observed. However, the formation of binary black holes includes many periods of mass loss, and the formation site itself could provide enough gas to produce an electromagnetic signal. The possible situations in which gas indeed is present at the merger site include the binary traveling through the
interstellar medium \citep{2019ApJ...884...22A}, the binary embedded within an active galactic nuclei disk \citep{2021ApJ...911..124L,2022MNRAS.517.1602L,2023ApJ...944...44K}, or the binary nested by the fallback supernova gas following black hole formation \citep{2016ApJ...821L..18P,2017ApJ...846L..15B,2018ApJ...862L...3S}. It has become increasingly evident that the most plausible LIGO progenitor channels are expected to lead to a black hole with a debris torus system. An important point is that the overall energetics of these various progenitors' avenues differ by many orders of magnitude, the spread reflecting the different masses left behind in the orbiting debris at the time of merger. Yet, a currently unexplored aspect of these progenitor systems is that the accretion luminosity at the time of merger is likely to be modulated by the tidal interaction from the companion. The results of the simulations presented in this study suggest that spiral waves are indeed a robust feature of accretion disks in binary black hole systems and that these spiral shocks can indeed
transport mass and angular momentum. Our goal in this work is to measure an effective viscosity due to spiral shocks excited in accretion disks by the tidal force of the black hole binary companion. 

To quantify the angular momentum transport by spiral shocks, we define the effective viscosity $\alpha_{\rm eff}=\dot{M}/3\pi\Sigma c_{\rm s}H$ following \citet{ju2016global}, where $\Sigma$ is the average gas surface density, $c_{\rm s}$ is the local isothermal sound speed, $H=c_{\rm s}/\Omega$ is the local scale height. The accretion rate $\dot{M}$ is the mass flux through each radius in the radial direction, including contributions from the Reynolds stress, the torque of the companion and the parameterized kinematic viscosity. 

In the simulations, we measure $\dot{M}=\int_{\phi}\rho v_{\rm r}dA$ and $\alpha_{\rm eff}$ by averaging corresponding quantities over $10^2$ orbital timescales at $R=0.02$, which is close to the inner boundary and approximates the accretion rate near the black hole. We also compared $\dot{M}$ at $R=0.025$ and $R=0.01$ and found no significant difference within the estimated $\dot{M}$ errors. We find $\alpha_{\rm eff}\approx 1.1\times10^{-2}$ for both \textsf{Am5} and \textsf{Am3}, which are orders of magnitude larger than the viscous stress. This suggests that spiral shock dissipation dominates the angular momentum transport in these disks, consistent with the findings reported in  Figure~\ref{fig:torques}.

\begin{figure}
    \centering
    \includegraphics[width=1.0\linewidth]{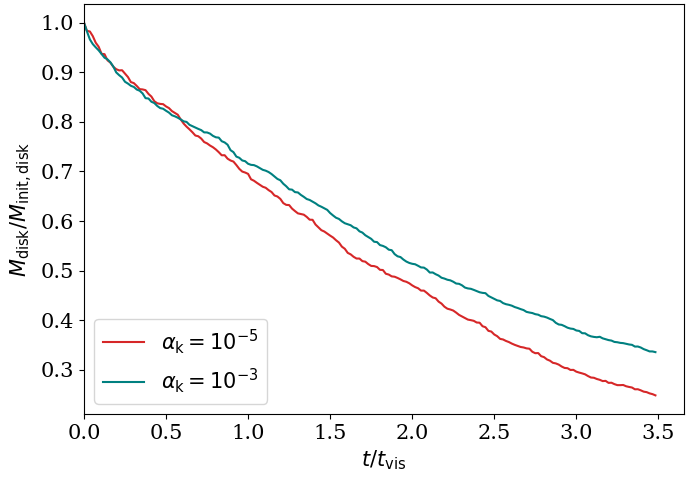}
    \caption{Disk draining in LIGO progenitors. Shown is the evolution of disk mass (spanning $0.02<R<0.25$) with the companion present, normalized to the initial disk mass at the disk draining phase. The {\it red} curve shows the disk mass for \textsf{Am5} with $\alphak=10^{-5}$, and the {\it green} curve shows disk mass for \textsf{Am3} (green curve) with $\alphak=10^{-3}$. The measured $\alpha_{\rm eff}\approx8.9\times10^{-3}\gg10^{-5}$ in \textsf{Am5}, yielding a viscous timescale $t_{\rm vis}\approx55.9$ that is roughly $3.6\times10^{3}$ orbital timescales. In \textsf{Am3} we measure $\alpha_{\rm eff}\approx1.3\times10^{-2}$ and $t_{\rm vis}\approx32.0$, which is roughly $2.1\times10^{3}$ orbital timescales. In the above plot, we scale the time to $t_{\rm vis}\approx55.9$, equivalent to effective viscosity $\alpha_{\rm eff}=1.0^{-2}$, roughly consistent with both \textsf{Am3} and \textsf{Am5}. }
    \label{fig:accretion}
\end{figure}

Figure~\ref{fig:accretion} shows the temporal evolution of the accretion disk mass ($0.02<R<0.25$) in the simulations. Varying the radial range from $0.02<R<0.3$ and $0.02<R<0.4$ did not lead to any marked differences in our results. In Figure~\ref{fig:accretion}, we normalize the time to the local viscous time measured at $R=0.02$. With the presence of companion and spiral shocks, \textsf{Am5} and \textsf{Am3} show significant mass loss, irrespective of the effective kinematic viscosity. For comparison, the mass loss in \textsf{Am5\_nc} and \textsf{Am3\_nc} is $\lesssim95\%$ of the initial disk mass at the beginning of the disk draining phase. The small amount of accretion is primarily due to the residual Reynold stresses that are generated from the disk-building phase. The residual spiral shocks mostly disappear after tens of orbital periods, and the disk relaxes to a smooth Keplerian disk.

With the fiducial scaling described in Section~\ref{sec:eq_scaling}, the initial disk mass in the above plot corresponds to $M_{\rm init}\approx 4\times10^{-3}M_{\odot}$. We also tested a set of similar simulations with two orders of magnitude lower disk mass $M_{\rm init}$, which are not shown in Figure~\ref{fig:accretion} as they show a consistent evolution. The initial disk masses we tested fall in between those observed above the neutrino cooling phase and below the photon Eddington accretion limit (Figure~\ref{fig:accretion_fallback}). Irrespective of the initial mass and viscous stress, we find that $M_{\rm disk}/M_{\rm init, disk}$ drains effectively within a few viscous timescales due to efficient angular momentum transport when a companion is present (Figure~\ref{fig:accretion}). We thus conclude that the binary tidal forces dominate the transport of mass and angular momentum within relic accretion disks, thus dimming the prospects for detectable electromagnetic counterparts to gravitational wave emissions from merging binary black holes. This, in turn, complicates attempts to study the host environments of these binary systems.\\ 

There are, however, a few caveats to the work presented here. For example, we do not include the role of magnetic fields in the simulations. The excitation of the magneto-rotation instability (MRI) is expected to contribute to the angular momentum transport and accretion, although most studies put the expected effective viscosity at values similar to those studied here. We thus expect that the inclusion of magnetic fields would not dramatically change our findings. What is more, the study presented in this {\it Letter} is motivated by the possibility that a weakly ionized disc might remain at the time of merger, which justifies the omission of magnetic fields in the simulations. Running our simulations in two dimensions also means that angular momentum transport is confined to the orbital plane. Such a configuration could overestimate the angular momentum transport in the radial direction relative to the vertical direction, which may affect the accretion rate \citep[e.g.,][]{2013A&A...556A.148P}. However, we anticipate that for a similar set-up in three-dimensional simulations, the effects of the companion's tidal force is also an important driver of angular momentum and mass transport and leading to some level of disk draining. Future exploration will help us determine how sensitively the spiral wave features depend on various disk properties. This could include varying Mach number as a function of scale height or radius, introducing a more physically-motivated cooling prescription, and exploring the effect of varying the mass ratio of the binary and the eccentricity of the companion's orbit.

\section*{Acknowledgements}
We thank A. Antoni, M. MacLeod, R. Rafikov, J. Stone, and Z. Zhu for fruitful discussions and guidance. We thank the referee for their thorough review, which significantly improved the quality of our manuscript. The UCSC team is supported in part by the Heising-Simons Foundation, the Vera Rubin Presidential Chair for Diversity at UCSC, and the National Science Foundation (AST-2307710, AST-2206243, AST-1911206, DGE-1842400, and AST-1852393). This work was supported by NASA Astrophysics Theory Program grant 80NSSC18K1018. The authors acknowledge use of the lux supercomputer at UC Santa Cruz, funded by NSF MRI grant AST 1828315. XH is supported by the Sherman Fairchild Postdoctoral Fellowship at the California Institute of Technology. SWD acknowledges funding from the Virginia Institute for Theoretical Astrophysics (VITA), supported by the College and Graduate School of Arts and Sciences at the University of Virginia. This work used the computational resources provided by the Advanced Research Computing Services (ARCS) at the University of Virginia. This work used Stampede 2 at Texas Advanced Computing Center through allocation AST150042 from the Advanced Cyberinfrastructure Coordination Ecosystem: Services $\&$ Support (ACCESS) program, which is supported by National Science Foundation grants 2138259, 2138286, 2138307, 2137603, and 2138296.  
\newpage 
\appendix

\section{The Linear Hydrodynamic Wave}\label{appendix:spiralshock}
Following \citet{ju2016global}, we fit the spiral shocks in the simulations by a linear wave dispersion relation. The results of this exercise are displayed in Figure~\ref{fig:spiralplots}. The phase of the waves takes the following form 
\begin{equation}
    \Phi_{m}=\int k(R)dR + m(\phi-\Omega_{\rm p}t),
\end{equation}
where $k$ and $m$ are the radial and azimuthal wavenumber \citep{1987gady.book.....B}. $\Omega_{\rm p}$ is the pattern speed and $\phi$ is the azimuthal angle. At a specific time $t$, the lines of constant-phase satisfy:
\begin{equation}
    \frac{d\phi}{dR}=-\frac{k}{m}
\end{equation}
The dispersion relation for hydrodynamic waves in a two-dimensional disk is
\begin{equation}
    m^{2}\left[\Omega(R)-\Omega_{\rm p}\right]^{2}=\kappa^{2}+c_{\rm s}^{2}k^{2}
\end{equation}
Here $\kappa$ is the epicyclic frequency, and $c_{\rm s}$ is the local sound speed. Assuming the disk is Keplerian, we have $\kappa=\Omega(R)\approx\sqrt{GM_{1}/R^{3}}$. We compare the time- and azimuthally-averaged radial disk angular momentum from simulations with the Keplerian profile in Figure~\ref{fig:spiralplots} and only found a noticeable deviation at the outer radius ($R \gtrsim 0.2$). Solving for $k$, the pitch angle of a trailing spiral arm is:
\begin{equation} 
    \frac{1}{R}\frac{dR}{d\phi}=\frac{d\log R}{d\phi}=-\frac{1}{ \mathcal{M}}\frac{1}{\{[1-\Omega_{\rm p}/\Omega(R)]^{2}-1/m^{2}\}^{1/2}}
    \label{eq:pitchangle}
\end{equation}
where $\mathcal{M}$ is the local Mach number, and assuming $\Omega_{\rm p}$ is the corotating angular speed.

\section{Angular Momentum Transport in Steady State Accretion Disks}\label{appendix:amtransfer}

Here, we briefly describe the angular momentum diagnostics used in this work. Our analysis follows \citet{ju2016global}, and we refer the reader to their paper for a more detailed derivation. An outline of the derivation of Equation~\ref{eq:per_ang} from Section~\ref{sec:am}  follows. First, the angular momentum conservation equation is integrated in the $z$- and $\phi$- directions using cylindrical coordinates. Then we use the mass conservation equation multiplied by $v_{\rm K}$ to eliminate canceling terms arising from steady Keplerian flow.

We write the $\phi$ component of velocity as $v_{\phi}=v_{\rm K}+\delta v_{\phi}$, where $v_{\rm K}$ is the Keplerian velocity, and $\delta v_{\phi}$ is the velocity perturbation. The equation for perturbed angular momentum can then be written as
\begin{equation}\label{eq:perturbedAMeq}
    \partial_{t}\langle \rho R\delta\vphi\rangle=-\langle R\rho\vr\rangle\frac{1}{R}\partial_{R}(R\vk)-\frac{1}{R}\partial_{R}(R^{2}\langle \rho\vr\delta\vphi\rangle)+\frac{1}{R}\partial_{R}(R^{2}\langle\rho\nu\Pi_{\rm R\phi}\rangle)+R\langle\rho a_{\rm ext,\phi}\rangle,
\end{equation}
where $\Pi_{\rm R\phi}$ is component of viscosity stress tensor. The second term on the right-hand side (RHS) originates from the $R\phi$-component of momentum tensor $M_{R\phi}=\rho v_{\rm R}v_{\phi}-B_{\rm R}B_{\phi}$. It is obtained by using the chain rule to rewrite $-\langle \frac{1}{R}\frac{\partial}{\partial R}(R^{2}\rho v_{\rm R}v_{\rm K})\rangle$. The notation $\langle X\rangle$ here represents the integral from $z_{\rm min}$ to $z_{\rm max}$ in vertical direction, and from $0$ to $2\pi$ in azimuthal direction within an annulus of size $\delta R$ at radius $R$. In the two-dimensional simulations, the vertical integration from $z_{\rm min}$ to $z_{\rm max}$ is simply the vertical direction domain range, and $\delta R$ is determined by local grid size in the R direction. 

The left-hand side in Equation~\ref{eq:perturbedAMeq} gives the rate of angular momentum variation ($\rm AM_{t}$), which is expected to vanish when the disk reaches steady state. On the right-hand side of Equation~\ref{eq:perturbedAMeq}, the different terms are related to angular momentum variations induced by the total mass accretion rate passing through a disk ring located at a radial distance $R$ ($\rm AM_{\dot{M}}$), by the radial gradient within the disk ring of the Reynolds stress ($\rm AM_{\rm FH}$) and the viscous stress ($AM_{\rm TM}$), and by the torque from the companion ($T$). In Figure~\ref{fig:torques}, we show these components for two of our simulations. In what follows, we describe how the time-integrated version of these terms are numerically computed.   

First, the left-hand-side of Equation~\ref{eq:perturbedAMeq} can be rewritten as
\begin{eqnarray}\label{eq:amdt}
 \text{AM}_{t}(R)&=\int_{t}\frac{\partial \langle \rho R\delta\vphi\rangle }{\partial t}\approx\bigg(\sum\limits_{k,j}\rho_{k,j,i}R_{i}[v_{i,k,j}-\vk(R_{i})]V_{k,j,i}\bigg)\bigg\rvert_{t_{1}}^{t_{2}},
\end{eqnarray}
where the subscripts $k,j,i$ label the cell indexes in the $z$, $\phi$ and $R$ directions respectively, and $V_{k,j,i}$ is the volume of each cell. On the right-hand-side of Equation~\ref{eq:perturbedAMeq}, the first term can be recast as
\begin{eqnarray}\label{eq:ammdot}
    \rm AM_{\dot{M}}&=&\frac{1}{R}\int_{t}-\langle R\rho\vr\rangle \partial_{R}(R\vk)\nonumber\\
    &\approx&-\sum\limits_{\delta t}\sum\limits_{k,j}R_{i}\rho_{\rm k,j,i}v_{\rm R,k,j,i}\left[\vk(R_{i+1/2})A_{R,i+1/2}-\vk(R_{i-1/2})A_{R,i-1/2}\right]\delta t,
\end{eqnarray}
where we approximate the mass flux $\rho_{\rm k,j,i}v_{\rm R,k,j,i}$ by the mass flux calculated by the Riemann solver at left interfaces in the radial direction. $A_{R}$ is the cell face area that norm to $R$-direction. The notation of $\sum\limits_{\delta t}$ represents that we approximate the integral as a sum over time steps.

The second term on the right-hand-side of Equation~\ref{eq:perturbedAMeq}, associated with the advection of angular momentum in the radial direction, can be recast  as
\begin{eqnarray}\label{eq:amth}
    \text{AM}_{\rm FH}&=&-\frac{1}{R}\int_{t}\langle \partial_{R}(R^{2}\rho\vr\delta\vphi)\rangle=-\frac{1}{R}\int_{t}\langle \partial_{R}(R^{2}T_{\rm H})\rangle
\end{eqnarray}
where $T_{\rm H}=\rho\vr\delta\vphi$ is the hydrodynamics Reynolds stress.  The viscosity term, $\text{AM}_{\rm TH}$, takes a similar form, except the Reynolds tensor is replaced by $\nu\Pi$. 
In Athena++, the second and third terms on the right-hand-side of Equation~\ref{eq:perturbedAMeq} are both included in the $\phi$-component of the momentum flux in $R$-direction, $\mathcal{F}(\rho\vphi, R)$, the subscript $i\pm1/2$ notes the variables calculated at the cell faces. The  total stress can therefore be calculated as
\begin{eqnarray}
    \text{AM}_{\rm FH}+\text{AM}_{\rm TM}&\approx&-\sum\limits_{\delta t}\sum\limits_{k,j}[R_{i+1/2}A_{R,i+1/2}\left[\mathcal{F}(\rho\vphi,R)_{k,j,i+1/2}-\vk(R_{i+1/2})\mathcal{F}(\rho,R)_{k,j,i+1/2}\right] \nonumber\\
    &-& R_{i-1/2}A_{R,i-1/2}\left[\mathcal{F}(\rho\vphi,R)_{k,j,i-1/2}-\vk(R_{i-1/2})\mathcal{F}(\rho,R)_{k,j,i-1/2}\right] \delta t
\end{eqnarray}

The last term on the right-hand-side of Equation~\ref{eq:perturbedAMeq}, associated with the contribution from external torque exerted by the companion, can be rewritten as
\begin{eqnarray}\label{eq:amtorque}
    \rm T(R)=\int_{t}\int_{V}R\rho a_{\rm ext}\approx\sum\limits_{\delta t}\sum\limits_{k,j}R_{i}\rho a_{\rm ext,k,j,i}V_{k,j,i}\delta t.
\end{eqnarray}

\newpage
\bibliographystyle{aasjournal}
\begin{scriptsize}
% \bibliography{Biblio}
\bibliography{bibfile}
\end{scriptsize}

\end{document}